\pdfoutput=1

\documentclass[11pt]{article}

\usepackage[]{EMNLP2023}

\usepackage{times}
\usepackage{latexsym}
\usepackage{graphicx}
\usepackage{amsmath}
\usepackage{amssymb}
\usepackage{float}
\usepackage{booktabs}
\usepackage{tabularx}
\usepackage{multirow}
\usepackage{array}
\usepackage{colortbl}
\usepackage{placeins}
\usepackage[moderate]{savetrees}

\usepackage[T1]{fontenc}

\usepackage[utf8]{inputenc}

\usepackage{microtype}

\usepackage{inconsolata}

\usepackage{color}
\usepackage{amsmath}

\newcommand{\JinaColbertTwo}{\href{https://huggingface.co/jinaai/jina-colbert-v2}{\texttt{Jina-ColBERT-v2}}}
\newcommand{\Colbert}{ColBERT}
\newcommand{\ColbertTwo}{ColBERTv2}
\newcommand{\XLMRoberta}{XLM-RoBERTa}
\newcommand{\JinaXLMRoberta}{Jina-XLM-RoBERTa}

\title{\JinaColbertTwo{}: A General-Purpose Multilingual Late Interaction Retriever}

\author{
    Rohan Jha$^{1}$\thanks{\quad Work done while at Jina AI.} \and
    Bo Wang$^2$ \and
    Michael G\"unther$^2$ \and Georgios Mastrapas$^2$ \and \\ {\bf Saba Sturua}$^2$ \and
    {\bf Isabelle Mohr}$^2$ \and {\bf Andreas Koukounas}$^2$ \and
    {\bf Mohammad Kalim Akram}$^2$ \and \\ {\bf Nan Wang}$^2$ \and {\bf Han Xiao}$^2$ \\
    $^1$The University of Texas at Austin, Austin, Texas, USA \\
    $^2$Jina AI GmbH, Prinzessinnenstr. 19-20, 10969 Berlin, Germany \\
    \texttt{research@jina.ai}
}

\begin{document}
\maketitle
\begin{abstract}

Multi-vector dense models, such as ColBERT, have proven highly effective in information retrieval.
ColBERT's late interaction scoring approximates the joint query-document attention seen in cross-encoders while maintaining inference efficiency closer to traditional dense retrieval models, thanks to its bi-encoder architecture and recent optimizations in indexing and search.
In this work we propose a number of incremental improvements to the ColBERT model architecture and training pipeline, using methods shown to work in the more mature single-vector embedding model training paradigm, particularly those that apply to heterogeneous multilingual data or boost efficiency with little tradeoff.
Our new model,  \JinaColbertTwo{}, demonstrates strong performance across a range of English and multilingual retrieval tasks.

\end{abstract}



\section{Introduction}
\label{sec:intro}

Neural retrieval has gained popularity in recent years following the arrival of capable pre-trained language models (PLMs) \citep{devlin_bert_2019, liu2019roberta, clark_electra_2020}. Two types of approaches have been employed to apply PLMs to retrieval.
Sparse neural retrieval systems, such as SPLADE \citep{formal_splade_2021}, represent texts as weighted bags of words that are interpreted as sparse high-dimensional vectors for maximum inner product search (MIPS).
Dense retrievers similarly encode queries and documents as \textit{dense} vectors, capturing relevance signals through spatial relationships extending beyond exact term matching.

Most dense retrievers encode a query or document as a single vector, commonly the result of mean-pooling or the [CLS]-embedding over the transformer’s final layer token embeddings.
In contrast, recent multi-vector retrievers like ColBERT \citep{khattab_colbert_2020} generalize this embedding process to maintain an embedding for each token, computing relevance scores as a function of the similarities of query and document tokens instead. To make the \Colbert{} usable in practice, the output dimensionality is restricted to be much smaller than the single-vector models.
This approach has the benefit of remaining compatible with much of the vector similarity infrastructure that makes single-vector methods efficient, but requires more space to store even a smaller embedding per token and compute at inference time to aggregate token interactions into a single score.
This late interaction over token embeddings achieves greater in-domain performance and tends to be more robust out-of-domain than single-vector similarity.
While ColBERTv2 is trained only on English MSMARCO triplets \citep{bajaj2016msmarco} and has a monolingual BERT backbone, making it incapable of multilingual retrieval, some previous works extend the model to multilingual retrieval.

ColBERT-XM \citep{louis_colbert-xm_2024} does this by using parameter extensions for each additional language, and \citep{lawrie_neural_2023} trains solely on machine-translated English MSMARCO data to get effective heterogeneous multilingual performance.
These approaches, however, come with trade-offs in terms of model usability and training data diversity. Other multilingual multi-vector models like BGE-M3 \citep{chen_bge_2024} produce extremely large token representations that limit their practical utility for first-stage retrieval.


In this work, we propose \JinaColbertTwo{}, which introduces an improved training recipe for \Colbert{} models with the following features:

\begin{description}
    \item{\textbf{Training with diverse weakly-supervised data:}} We additionally pretrain our modified PLM with rotary position embedding and train on large-scale unlabeled text pairs from various corpus with a weakly-supervised single-vector contrastive objective. A second-stage of ColBERT finetuning with labeled triplet data and supervised distillation is used to further boost its performance.\\
    \item{\textbf{General multilingual performance:}} We train with data from a variety of high- and low-resource languages using both labeled and unlabeled data, including human- and machine-translated training data, and show that this improves even out-of-domain multilingual performance.\\
    \item{\textbf{Inference-agnostic efficiency:}} We introduce multiple sizes of linear projection heads, jointly trained using the non-weight tying variant of Matryoshka Representation Loss \cite{kusupati_2022_matryoshka}, enabling the selection of token embedding size at inference time with minimal performance degradation. We demonstrate that reducing the embedding dimensionality in half from 128 to 64 yields only a minor performance tradeoff. Additionally, our flash-attention optimized backbone, \JinaXLMRoberta{}\, provides further free performance improvement during inference.\\
\end{description}

Our experimental results show competitive retrieval performance across both English and multilingual benchmarks. We also present controlled experiments demonstrating the benefits, or lack thereof, of the training modifications we consider in developing our training recipe.

\section{Related Work}
In this section, we discuss related work in single- and multi-vector retrieval, as well as the non-English late-interaction retrievers from which our training recipe draws inspiration.
\label{sec:related_work}
\subsection{Single-Vector Retrieval}

Single-vector encoder models have demonstrated their potential as general-purpose embedding models across a number of downstream tasks \citep{muennighoff-etal-2023-mteb}.
When used in a bi-encoder retrieval model, they asymmetrically encode queries and documents as separate dense vectors, and measure their pairwise relevance as the cosine similarity between the vectors.
Owing to their strong in-domain performance and straightforward inference scheme, there has been a growing focus on improving their training.
Studies demonstrate that large-scale unsupervised pair training utilizing in-batch negatives, followed by a small-scale triplet finetuning stage, significantly improves performance compared to a dense retriever trained solely on triplet data \citep{li2023generaltextembeddingsmultistage, gunther2023jina}.
Other works have incorporated asymmetric task-specific instructions for queries and documents to further enhance performance \citep{wang2024multilingual} and demonstrated the efficacy of using synthetically generated training data, including using diverse task instructions and machine translations, to further improve model representations. \citep{wang2023improving, lee_gecko_2024}

\subsection{Multi-Vector Retrieval}

Multi-vector retrievers like \Colbert{}\ also employ a bi-encoder structure, but queries and passages are represented by a collection of smaller token embeddings rather than one large vector.
As such, \ColbertTwo{}'s training uses many of the same techniques as state-of-the-art single-vector models: cross-encoder distillation, multiple negatives per query, and self-mined hard negatives.
Recent models have continued to improve on this training recipe, particularly for multilingual or non-English training. BGE-M3 \citep{chen_bge_2024} adopts the two-stage pairs-to-triplets training pipeline, and does self-knowledge distillation, treating the combination of its sparse, dense, and multi-vector scores as the teacher score.

\subsection{Multilingual Retrieval}
Owing to the quality of English-based pre-trained models (BERT) and annotated data (MSMARCO), many advances in neural retrieval have been applied first to the monolingual English setting \cite{karpukhin_dense_2020, xiong_approximate_2020, khattab_colbert_2020}.
Researchers, however, have also made advances in non-English capabilities.

On the modeling front, multilingual PLMs like mBERT \citep{devlin_bert_2019} and later \XLMRoberta{}\ \citep{conneau_unsupervised_2020} have expanded pre-training to include text in up to 100 languages, including in cross-language contexts.
For multilingual retrieval data, there are two approaches: natural and translated.
Datasets like Mr-Tydi and MIRACL \citep{mrtydi, zhang2023miracl} are built from human-generated and annotated queries, whereas mMARCO \citep{bonifacio_mmarco_2022} is a collection of machine-translated copies of MSMARCO which inherit their judgments from the original dataset.
The former method tends to be of higher quality and lacks the subtle distributional/idiomatic errors, dubbed "translationese", that the latter sometimes exhibits.
Naturally, however, human generation costs more per example.

Recent multi-vector work has also proposed further modifications along the dimensions of architecture and data.
ColBERT-XM \cite{louis_colbert-xm_2024} addresses the so-called \textit{curse of multilinguality} \citep{conneau_unsupervised_2020}, the performance degradation of models pre-trained on too many tasks, with shared- and per-language parameters that allow for more robust zero-shot language transfer and post-hoc language extension.
On the data approach, ColBERT-X \citep{ecir2022colbert-x, lawrie_neural_2023, yang_distillation_2024} uses language-mixed batches of machine-translated English data, and BGE-M3 \citep{chen_bge_2024} curates unsupervised and high-quality supervised corpora of diverse multilingual training data.

\section{Training Overview}
\JinaColbertTwo{}'s training paradigm has three parts:
\begin{enumerate}
    \item \textbf{Modified Encoder Architecture}: We use a modified encoder backbone, derived from \XLMRoberta{}\ with improvements made to its architecture and pre-training regime.
    We further extend \Colbert{}'s linear projection head by jointly training a collection of different-size heads for embedding size reduction.

    \item \textbf{Pair Training}: To learn from the semantic structure of large quantities of diverse data in many languages, we first train our encoder model on weakly supervised text pairs from a variety of embedding datasets.

    \item \textbf{Triplet Training}: Our model is further finetuned using retrieval examples in many languages with both positives and hard negatives, supervised by a highly-capable multilingual cross-encoder.
\end{enumerate}

The following sections describe our experiments on these three components of training \JinaColbertTwo{}.
\section{Architecture}
\subsection{Backbone Improvements}

Following many prior single- and multi-vector multilingual training efforts, we adopted \XLMRoberta{} as our backbone model due to its strong performance across various downstream tasks \cite{ecir2022colbert-x, louis_colbert-xm_2024, chen_bge_2024}. To improve the efficiency, we enhance the \XLMRoberta{} architecture with flash attention \citep{dao2023flashattention2}.

We replace the absolute positional embeddings with rotary positional embeddings (RoPE, \citet{su2023roformerenhancedtransformerrotary}), which are empirically understood to be better. They also have the advantage of supporting context lengths far longer than 512 tokens, although we do not explicitly focus on long-context in this work.
To warm up its new positional embeddings, we continued pre-training the modified backbone
with the same masked language modeling objective for 160,000 steps on the RefinedWeb dataset \cite{penedo2023refinedweb}, a modern, high-quality corpus, under the masked language modeling objective.
During this pre-training phase, we set the maximum sequence length to 8,192 tokens with a rotary base of 10,000 and employed whole-word-masking \cite{devlin_bert_2019}, masking out 30\% of the tokens. We call this modified language model \JinaXLMRoberta{}.


\subsection{Multiple Linear Heads}
To reduce index sizes, \Colbert{} includes a linear head that projects its token embeddings from the hidden dimension of its language model down to a lower dimension ($768 \to 128$).
As a notable exception, BGE-M3's multi-vector retrieval does not take this step, keeping its token embeddings at a full 1024 dimensions.

\begin{figure}[h]
    \centering
    \includegraphics[width=\linewidth]{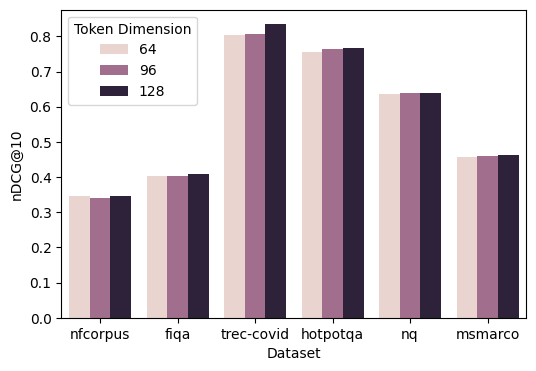}
    \caption{nDCG@10 scores for BEIR datasets using 64-, 96-, and 128-dimension linear projection heads for token embeddings.}
    \label{fig:mrl_scores}
\end{figure}

We jointly train six linear heads with dimensions $d \in \{64, 96, 128, 256, 512, 768\}$ using Matryoshka Representation Loss (MRL, \citet{kusupati_2022_matryoshka}).
This allows users to choose greater or lesser space efficiency, with an associated performance trade-off.
Figure \ref{fig:mrl_scores} quantifies this tradeoff, showing the strong performance preservation of our reduced-dimension linear heads. Halving the token dimension ($128 \to 64$) only causes its nDCG@10 to drop by 0.01 (1.59\%).
We unfortunately find that MRL's weight-tying efficient variant (MRL-E), where losses are computed on \textit{truncations} of the same token vector does not preserve performance well, which we hypothesize is a consequence of the already-low projected dimension of the original \Colbert{}\ formulation.

\section{Pair Training}
\label{sec:pair_training}

To leverage an abundance of text pairs with varying richness of semantic structure, we draw inspiration from common practices in single-vector embedding model training and begin with training on these text pairs, focusing on optimizing the embedding model's performance on general semantic similarity and relatedness tasks.
This weakly-supervised stage is in contrast to previous \Colbert{} works, which typically start directly from a PLM like BERT with triplet training on 32-way or 64-way retrieval triplets consisting of a query, a positive passage, and multiple mined negatives.

\subsection{Data Composition}
\label{sec:data_composition}

Our pair training data consists of a broad range of weakly supervised datasets harvested from the web.
We adjusted sampling rates across different languages and domains based on intuition, resulting in a set of 450 million weakly supervised, semantically related sentence pairs, question-answer pairs, and query-document pairs.
Of these 450 million pairs, 50.0\% are in English.
Our non-English pair-wise datasets contain a diverse collection of 29 major languages, including 3.0\% code data, with 4.3\% representing cross-lingual data.

\subsection{Contrastive Loss}
\label{sec:contrastive_loss}

We utilize the same \textit{single-vector} pair-training loss function as described in \cite{gunther2023jina}.
Due to the often symmetric nature of our text pairs, the loss is calculated in both directions.
During the pair training stage, we set the temperature $\tau=0.02$ and used a peak learning rate of $5 \times 10^{-5}$ with a warm-up period of 1,000 steps.
The model was trained using the Adam optimizer for 100,000 steps with a global batch size of 16,384.

\section{Triplet Training}
\subsection{Data Composition}
Our triplet dataset consists of 1) high-quality, human-annotated research datasets such as MSMARCO, DuReader, and MIRACL \citep{bajaj2016msmarco, he2018dureaderchinesemachinereading, zhang2023miracl} with diversely mined negatives 2) high-quality datasets like MSMARCO and NQ translated from English into Chinese, French, German, Japanese, Russian and Spanish, following our previous work \cite{mohr2024multi} and 3) synthetically generated datasets to address common failure modes of dense vector models such as negation and to cover niche domains like legal IR.

The triplet dataset covers 14 widely used languages, with a strong emphasis on Arabic, Chinese, English, French, German, Japanese, Russian, and Spanish. We sample the datasets to create a language distribution similar to that used in pair training.
English accounts for 45.9\% of the triplets, with 52.1\% roughly evenly split between the mentioned high-resource non-English languages and a small 2.0\% share for lower-resource languages.

Notably, owing to the limitations of our various sources of data, we train on triplets with only 7 negatives per example, in contrast to the 32- or 64-way triplets of \ColbertTwo{}.

\subsection{Supervision Loss}

Following \ColbertTwo{}, we finetune our pair-trained checkpoint on samples with hard negatives using a KL divergence loss function to distill soft labels from the teacher model.
For the teacher model, we use \texttt{jina-reranker-v2-base-multilingual}\footnote{\url{https://huggingface.co/jinaai/jina-reranker-v2-base-multilingual}}, a highly capable multilingual cross encoder.

This stage trains for 100,000 steps with a batch size of 32 and a cosine decay learning rate schedule with 5\% warm-up that peaks at $1 \times 10^{-5}$. We use pure BFLOAT-16 precision, and apply magnitude-based gradient clipping with a threshold of $1$ for stability.

\section{Results}

\begin{table*}[htbp]
    \centering
    \small


\centering
\begin{tabular}{c|r|rrrrrrrrrrrrr} 
\toprule
\textbf{BEIR}                    & \multicolumn{1}{l|}{\textbf{avg}} & \multicolumn{1}{l}{\textbf{nf}} & \multicolumn{1}{l}{\textbf{fi}} & \multicolumn{1}{l}{\textbf{tc}} & \multicolumn{1}{l}{\textbf{ar}} & \multicolumn{1}{l}{\textbf{qu}} & \multicolumn{1}{l}{\textbf{sd}} & \multicolumn{1}{l}{\textbf{sf}} & \multicolumn{1}{l}{\textbf{to}} & \multicolumn{1}{l}{\textbf{db}} & \multicolumn{1}{l}{\textbf{fe}} & \multicolumn{1}{l}{\textbf{cf}} & \multicolumn{1}{l}{\textbf{hp}} & \multicolumn{1}{l}{\textbf{nq}}  \\ 
\midrule
BM25                             & 44.0                              & 32.5                            & 23.6                            & 65.6                            & 31.5                            & 78.9                            & 15.8                            & 66.5                            & \textbf{36.7}                   & 31.3                            & 75.3                            & 21.3                            & 60.3                            & 32.9                             \\
ColBERTv2                        & 49.6                              & 33.7                            & 35.4                            & 72.6                            & 46.5                            & 85.5                            & 15.4                            & 68.9                            & 26.0                            & 45.2                            & 78.5                            & 17.6                            & 67.5                            & 52.4                             \\
\multicolumn{1}{l|}{answerai-v1} & \textbf{55.7}                     & \textbf{37.3}                   & \textbf{41.2}                   & \textbf{84.6}                   & \textbf{50.1}                   & 87.7                            & 18.4                            & \textbf{74.8}                   & 25.7                            & 45.6                            & \textbf{91.0}                   & \textbf{33.1}                   & 76.1                            & 59.1                             \\ 
\midrule
Ours                             & 53.1                              & 34.6                            & 40.8                            & 83.4                            & 36.6                            & \textbf{88.7}                   & \textbf{18.6}                   & 67.8                            & 27.4                            & \textbf{47.1}                   & 80.5                            & 23.9                            & \textbf{76.6}                   & \textbf{64.0}                    \\
\bottomrule
\end{tabular}

    \caption{Comparison of nDCG@10 scores between BM25, ColBERTv2, answer-colbert-small and Jina-ColBERT-v1 and \JinaColbertTwo{} on the BEIR test set.  \textbf{nf} for NFCorpus, \textbf{fi} for FIQA (Fact In Question Answering), \textbf{tc} for TREC-COVID (Text Retrieval Conference COVID), \textbf{ar} for Arguana, \textbf{qu} for Quora, \textbf{sd} for SciDocs, \textbf{sf} for SciFact, \textbf{to} for Webis-Touche, \textbf{db} for DBpedia-Entity, \textbf{fe} for FEVER (Fact Extraction and Verification), \textbf{cf} for Climate-FEVER, \textbf{hp} for HotpotQA, and \textbf{nq} for Natural Questions}
    \label{tab:beir_final}
\end{table*}

We evaluate \JinaColbertTwo{}\ on four widely used benchmarks, BEIR, LoTTE, and MIRACL and mMARCO.
For general English performance, we use the same subset of 14 retrieval and text-similarity tasks from the BEIR benchmark as in \citet{santhanam_colbertv2_2022}. Additionally, we assess performance on the LoTTE benchmark, which focuses on long-tail queries, and the MIRACL and mMARCO benchmarks \cite{zhang2023miracl, bonifacio_mmarco_2022}, which assess non-English retrieval performance.
We report nDCG@10 for the BEIR and MIRACL collections, MRR@10 for mMARCO, and Success@5 for LoTTE.
Scores are reported on the test split for BEIR, development split for MIRACL and mMARCO, and search test split for LoTTE. We use the same maximum query/document lengths as reported in \citet{santhanam_colbertv2_2022}, and use the default (32/300) for MIRACL and mMARCO.

\begin{table}[htbp]
\small
\setlength{\tabcolsep}{3pt}
\centering
\begin{tabular}{r|c|ccccc}
\toprule
\textbf{LoTTE} & \textbf{avg}  & \textbf{Life.} & \textbf{Rec.} & \textbf{Wri.} & \textbf{Sci.} & \textbf{Tech.}  \\
\midrule
BM25           & 67.8          & 80.2           & 68.5          & 74.7          & 53.6          & 61.9            \\
ColBERTv2      & 72.0          & 84.7           & 72.3          & 80.1          & 56.7          & 66.1            \\
\midrule
Ours           & \textbf{76.4} & \textbf{87.0}  & \textbf{77.6} & \textbf{83.8} & \textbf{60.5} & \textbf{73.0}   \\
\bottomrule
\end{tabular}
\caption{Comparison of Success@5 of various models across different LoTTE search query subsets.}
\label{tab:lotte_comparison}
\end{table}

Table \ref{tab:beir_final} shows \JinaColbertTwo{}'s strong English performance compared to \ColbertTwo{}, while still trailing the monolingual answerai-colbert-small-v1.
Notably, however, we perform well below \ColbertTwo{} on ArguAna (\textbf{ar}), which we might attribute to either its unusual task: \textit{counterargument retrieval} being at odds with our retrieval-heavy triplet training data distribution, or as an indication of the limitation of our stronger augmentation attention (discussed in Section \ref{sec:query_augmentation_attention}) when applied to much longer (300 token) queries.
Similarly for LoTTE, we see in Table \ref{tab:lotte_comparison} an improvement over \ColbertTwo{}.

\begin{table*}[htbp]
\centering
\small
\setlength{\tabcolsep}{2pt}
\begin{tabular}{r|c|cccccccccccccccccc}
\toprule
\textbf{MIRACL} & \textbf{avg}  & \textbf{ar}   & \textbf{bn}   & \textbf{de}   & \textbf{es}   & \textbf{en}   & \textbf{fa}   & \textbf{fi}   & \textbf{fr}   & \textbf{hi}   & \textbf{id}   & \textbf{ja}   & \textbf{ko}   & \textbf{ru}   & \textbf{sw}   & \textbf{te}   & \textbf{th}   & \textbf{yo}   & \textbf{zh}    \\
\midrule
BM25            & 38.5          & 48.1          & 50.8          & 22.6          & 31.9          & 35.1          & 33.3          & 55.1          & 18.3          & 45.8          & 44.9          & 36.9          & 41.9          & 33.4          & 38.3          & 49.4          & 48.4          & 40.6          & 18.0           \\
mDPR-ZS         & 41.8          & 49.9          & 44.3          & 49.0          & 47.8          & 39.4          & 48.0          & 47.2          & 43.5          & 38.3          & 27.2          & 43.9          & 41.9          & 40.7          & 29.9          & 35.6          & 35.8          & 39.6          & 51.2           \\
mDPR-FT         & \textbf{62.7} & 72.5          & 68.4          & -             & 48.8          & 56.5          & \textbf{59.3} & 71.4          & \textbf{58.9} & 51.6          & 49.6          & \textbf{64.2} & 59.0          & 59.7          & \textbf{68.5} & \textbf{80.4} & 69.5          & -             & \textbf{65.0}  \\
\midrule
Ours            & 62.3          & \textbf{75.3} & \textbf{75.0} & \textbf{50.4} & \textbf{53.8} & \textbf{57.0} & 56.3          & \textbf{74.0} & 54.1          & \textbf{60.0} & \textbf{54.7} & 63.2          & \textbf{67.1} & \textbf{64.3} & 49.9          & 74.2          & \textbf{77.2} & \textbf{62.3} & 52.3        \\
\bottomrule
\end{tabular}

\caption{Comparison of nDCG@10 scores for BM25, mDPR-ZeroShot (ZS), mDPR-FineTuned (FT), and \JinaColbertTwo{} models on the MIRACL dev set across various languages.}
\label{tab:miracl_final}
\end{table*}

Table \ref{tab:miracl_final} compares \JinaColbertTwo{} to BM25, mDPR, and BGE-M3. While we handily outperform BM25 and zero-shot mDPR \citep{zhang2023miracl} as expected, our model is slightly outperformed by the finetuned mDPR \citep{zhang_2023_towards}. For context, each mDPR-FT is only tuned on one language, rather than many like ours which may suffer to some extent from the \textit{curse of multilinguality}.

\begin{table*}[htbp]
\centering
\small
\setlength{\tabcolsep}{3pt}
\begin{tabular}{r|c|ccccccccccccccc}
\toprule
\textbf{mMARCO} & \textbf{avg}  & \textbf{ar}   & \textbf{de}   & \textbf{nl}   & \textbf{es}   & \textbf{fr}   & \textbf{hi}   & \textbf{id}   & \textbf{it}   & \textbf{ja}   & \textbf{pt}   & \textbf{ru}   & \textbf{vi}   & \textbf{zh}    \\
\midrule
BM-25           & 13.9          & 11.1          & 13.6          & 14.0          & 15.8          & 15.5          & 13.4          & 14.9          & 15.3          & 14.1          & 15.2          & 12.4          & 13.6          & 11.6           \\
ColBERT-XM      & 25.4          & 19.5          & 27.0          & 27.5          & 28.5          & 26.9          & 23.8          & 26.3          & 26.5          & 24.1          & 27.6          & 25.1          & 22.6          & 24.6           \\
\midrule
Ours            & \textbf{31.3} & \textbf{27.2} & \textbf{33.1} & \textbf{33.0} & \textbf{34.1} & \textbf{33.5} & \textbf{30.9} & \textbf{31.9} & \textbf{33.7} & \textbf{27.6} & \textbf{33.7} & \textbf{29.8} & \textbf{28.7} & \textbf{30.2}
 \\
\bottomrule
\end{tabular}
\caption{Comparison of mRR@10 scores between BM25, ColBERT-XM and \JinaColbertTwo{} models on the mMARCO dev set across various languages.}
\label{tab:mmarco_final}
\end{table*}

Finally, comparing against ColBERT-XM's zero-shot evaluation on mMARCO in Table \ref{tab:mmarco_final}, we see a strong improvement across the board, including on languages whose mMARCO training set does not occur in our pair or triplet training data (dt, hi, id, it, pt, vi).

\section{Ablation Studies}
In this section we present short ablation studies on modifications to three various aspects of ColBERT modeling and training.
\subsection{Efficient Evaluation}
\label{sec:eval_method}
Due to the compute and time costs of indexing corpora containing tens of millions of documents, evaluating every model checkpoint and ablation on every task is not feasible.
Therefore, we follow recent works  \citep{clavie_jacolbertv25_2024, merrick2024arcticembedscalableefficientaccurate} by comparing models' quality on smaller sampled-corpus versions of HotpotQA, NQ, MS MARCO, and MIRACL (Chinese, French, German, Japanese, Spanish).
These sampled corpora are constructed by combining the top 250 BM25-retrieved\footnote{We use the standard pre-built Lucene indices in Pyserini \citep{Lin_etal_SIGIR2021_Pyserini} for MIRACL found at \url{https://github.com/castorini/pyserini}, and use BM25s \citep{bm25s} for BEIR.} passages with all judged passages. We observe good agreement between the sampled-corpus evaluation scores and the full-fidelity ones when used to make binary or ranking-based model comparisons, but we leave a more rigorous analysis of this observation to future work.
We only use the sampled corpora for ablation studies. For the final model, we evaluate on the full version of every dataset.

\subsection{Task Instructions}
\label{sec:instruciton}
Inspired by the use of instruction prefixes in single-vector works like \citet{su2022one}, we experimented with adding task-specific natural language instructions for retrieval (RET), and question answering (QA), and semantic text similarity (STS).
However, results in Table \ref{tab:instructions} show a generally negative effect across most BEIR datasets. We hypothesize that this is because instructions are not well-suited for late interaction models, which operate at the token level.
Any embedding conditioning that the instructions might provide likely becomes less effective when aggregated at the token similarity level. Furthermore, these instructions occupy valuable space within the system's fixed token capacity.

\begin{table*}[htbp]
    \centering
    \small

\begin{tabular}{rccccccccccccccc}
\toprule
& \multicolumn{8}{c}{\textbf{RET}} & \multicolumn{3}{c}{\textbf{QA}} & \multicolumn{3}{c}{\textbf{STS}} \\
\cmidrule(lr){2-9}\cmidrule(lr){10-12}\cmidrule(lr){13-15}
& \textbf{nf} & \textbf{tc} & \textbf{sf} & \textbf{to} & \textbf{db} & \textbf{fe} & \textbf{cf} & \textbf{ms}* & \textbf{fq} & \textbf{hp}* & \textbf{nq}* & \textbf{ar}  & \textbf{qu} & \textbf{sd} \\
\midrule
Mark. & 32.4 & 59.3 & \textbf{67.9} & \textbf{19.3} & \textbf{35.3} & \textbf{67.1} & \textbf{18.3} & \textbf{34.4} & \textbf{37.5} & \textbf{25.9} & 40.8 & \textbf{37.5} & \textbf{86.1} & \textbf{18.4} \\
Inst. & \textbf{32.9} & \textbf{63.2} & 67.5 & 18.8 & 33.9 & 64.4 & 16.7 & 34.0 & 37.1 & 24.9 & \textbf{42.9} & 34.2 & 86.0 & 17.9 \\
\bottomrule
\end{tabular}

    \caption{nDCG@10 scores on BEIR datasets, grouped by task type (retrieval, question answering, and semantic text similarity) when using natural language instructions versus query/document marker tokens (default). Datasets marked with a * use the BM25-sampled corpus technique discussed in Section \ref{sec:eval_method}.}
    \label{tab:instructions}
\end{table*}

\subsection{Score Normalization}

\begin{table*}[htbp]
    \centering
    \small
\begin{tabular}{rccccccccc}
\toprule
& \multicolumn{4}{c}{\textbf{BEIR}} & \multicolumn{5}{c}{\textbf{MIRACL}} \\
\cmidrule(lr){2-5}\cmidrule(lr){6-10}
 & \textbf{tc} & \textbf{hp} & \textbf{nq} & \textbf{ms} & \textbf{de} & \textbf{es} & \textbf{fr} & \textbf{ja} & \textbf{zh} \\
\midrule

Baseline & 78.7 & \textbf{36.6} & \textbf{58.0} & \textbf{45.4} & 57.3 & \textbf{40.6} & 50.7 & \textbf{63.4} & \textbf{63.2}\\
+ Score Norm. & \textbf{80.1} & 36.4 & 56.6 & 45.1 & \textbf{57.7} & 39.3 & \textbf{51.3} & 61.8 & 62.5\\
\bottomrule
\end{tabular}
    \caption{nDCG@10 scores with and without score normalization on a retrieval-oriented subset of BEIR and MIRACL tasks. Results are performed on the BM25-sampled versions of all datasets presented except TREC-COVID (\textbf{tc}).}
    \label{tab:minmax_ablation_table}
\end{table*}

Recently, \citet{clavie_jacolbertv25_2024} applied min-max normalization to both the student and teacher scores before computing the KL loss.
This adjustment brings the score distributions of the \Colbert{} model and its CE teacher into closer alignment, as the original score distribution for \Colbert{} theoretically ranges from zero to the number of query tokens, and is model-dependent for the teacher CE.
Our experiment presented in Table \ref{tab:minmax_ablation_table}, however, shows this method to have inconclusive benefit to nDCG@10 on the BEIR and MIRACL datasets when applied to our model. We consider this result to be understandable given \citet{clavie_jacolbertv25_2024}'s very small observed effect.

\subsection{Query Augmentation Attention}
\label{sec:query_augmentation_attention}

\begin{table*}[htbp]
    \centering
    \small
\begin{tabular}{rccccccccc}
\toprule
& \multicolumn{4}{c}{\textbf{BEIR}} & \multicolumn{5}{c}{\textbf{MIRACL}} \\
\cmidrule(lr){2-5}\cmidrule(lr){6-10}
& \textbf{tc} & \textbf{hp} & \textbf{nq} & \textbf{ms} & \textbf{de} & \textbf{es} & \textbf{fr} & \textbf{ja} & \textbf{zh} \\
\midrule

Baseline & 77.2 & 70.4 & 54.6 & 37.6 & 33.3 & 40.3 & 35.9 & 54.9 & 34.4\\
+ \texttt{[MASK]} attn. & \textbf{80.2} & \textbf{71.5} & \textbf{58.8} & \textbf{44.3} & \textbf{45.6} & \textbf{49.8} & \textbf{44.8} & \textbf{58.8} & \textbf{52.9} \\
\bottomrule
\end{tabular}
    \caption{nDCG@10 scores with and without query augmentation \texttt{[MASK]} token attention on a retrieval-oriented subset of BEIR and MIRACL tasks. Results report full-fidelity scores.}
    \label{tab:mask_ablation_table}
\end{table*}

An important feature of ColBERT's implementation is its query augmentation mechanism. By padding queries with \texttt{[MASK]} tokens to a uniform length, ColBERT uses BERT's masked language modeling ability to produce additional soft term embeddings which interact with document token embeddings during MaxSim scoring.
However, prior \Colbert{}\ models do not modify the attention mask to allow query tokens to attend to the mask tokens, which some hypothesize might harm generalization by making this augmentation feature too integral to the embedding process.
Our controlled triplet training experiment in Table \ref{tab:mask_ablation_table}, however, demonstrates a positive effect across a variety of tasks, with particular benefit to non-English tasks in MIRACL. We therefore allow this attention in our training and inference.



\section{Conclusion}
This work presents \JinaColbertTwo{}, a capable multilingual ColBERT model that is the result of improvements to its architecture and training process. We implement modifications to the model architecture that yield efficiency gains with effectively no downside, and subsequently train it on a heterogeneous mix of data of varying tasks, languages, and supervision structures in order to bolster its performance as a general purpose retriever.
Our ablation experiments demonstrate the sensitivity of ColBERT to modifications to its representations.

We hope that our work will support future multilingual ColBERT development, and prompt further exploration into the properties and optimal configuration of its query augmentation mechanism. We are also encouraged by the many inference-only optimization works on ColBERT representations, and suggest further effort be invested in tying these methods more closely with the models training objective.

\section{Acknowledgement}
We thank Qi Liu and Jiaxin Mao from Renmin University of China for the contributions to \texttt{Jina-ColBERT-v1} and offer the insights about MRL over MRL-E for ColBERT models.

\bibliography{references, zotero_8_6}

\begin{thebibliography}{36}
\expandafter\ifx\csname natexlab\endcsname\relax\def\natexlab#1{#1}\fi

\bibitem[{Bajaj et~al.(2016)Bajaj, Campos, Craswell, Deng, Gao, Liu, Majumder, McNamara, Mitra, Nguyen, Rosenberg, Song, Stoica, Tiwary, and Wang}]{bajaj2016msmarco}
Payal Bajaj, Daniel Campos, Nick Craswell, Li~Deng, Jianfeng Gao, Xiaodong Liu, Rangan Majumder, Andrew McNamara, Bhaskar Mitra, Tri Nguyen, Mir Rosenberg, Xia Song, Alina Stoica, Saurabh Tiwary, and Tong Wang. 2016.
\newblock \href {http://arxiv.org/abs/1611.09268} {Ms marco: A human generated machine reading comprehension dataset}.

\bibitem[{Bonifacio et~al.(2022)Bonifacio, Jeronymo, Abonizio, Campiotti, Fadaee, Lotufo, and Nogueira}]{bonifacio_mmarco_2022}
Luiz Bonifacio, Vitor Jeronymo, Hugo~Queiroz Abonizio, Israel Campiotti, Marzieh Fadaee, Roberto Lotufo, and Rodrigo Nogueira. 2022.
\newblock \href {https://doi.org/10.48550/arXiv.2108.13897} {{mMARCO}: {A} {Multilingual} {Version} of the {MS} {MARCO} {Passage} {Ranking} {Dataset}}.
\newblock ArXiv:2108.13897 [cs].

\bibitem[{Chen et~al.(2024)Chen, Xiao, Zhang, Luo, Lian, and Liu}]{chen_bge_2024}
Jianlv Chen, Shitao Xiao, Peitian Zhang, Kun Luo, Defu Lian, and Zheng Liu. 2024.
\newblock \href {https://doi.org/10.48550/arXiv.2402.03216} {{BGE} {M3}-{Embedding}: {Multi}-{Lingual}, {Multi}-{Functionality}, {Multi}-{Granularity} {Text} {Embeddings} {Through} {Self}-{Knowledge} {Distillation}}.
\newblock ArXiv:2402.03216 [cs].

\bibitem[{Clark et~al.(2020)Clark, Luong, Le, and Manning}]{clark_electra_2020}
Kevin Clark, Minh-Thang Luong, Quoc~V. Le, and Christopher~D. Manning. 2020.
\newblock \href {http://arxiv.org/abs/2003.10555} {{ELECTRA}: {Pre}-training {Text} {Encoders} as {Discriminators} {Rather} {Than} {Generators}}.
\newblock ArXiv:2003.10555 [cs].

\bibitem[{Clavié(2024)}]{clavie_jacolbertv25_2024}
Benjamin Clavié. 2024.
\newblock \href {http://arxiv.org/abs/2407.20750} {{JaColBERTv2}.5: {Optimising} {Multi}-{Vector} {Retrievers} to {Create} {State}-of-the-{Art} {Japanese} {Retrievers} with {Constrained} {Resources}}.
\newblock ArXiv:2407.20750 [cs].

\bibitem[{Conneau et~al.(2020)Conneau, Khandelwal, Goyal, Chaudhary, Wenzek, Guzmán, Grave, Ott, Zettlemoyer, and Stoyanov}]{conneau_unsupervised_2020}
Alexis Conneau, Kartikay Khandelwal, Naman Goyal, Vishrav Chaudhary, Guillaume Wenzek, Francisco Guzmán, Edouard Grave, Myle Ott, Luke Zettlemoyer, and Veselin Stoyanov. 2020.
\newblock \href {https://doi.org/10.48550/arXiv.1911.02116} {Unsupervised {Cross}-lingual {Representation} {Learning} at {Scale}}.
\newblock ArXiv:1911.02116 [cs].

\bibitem[{Dao(2024)}]{dao2023flashattention2}
Tri Dao. 2024.
\newblock Flash{A}ttention-2: Faster attention with better parallelism and work partitioning.
\newblock In \emph{International Conference on Learning Representations (ICLR)}.

\bibitem[{Devlin et~al.(2019)Devlin, Chang, Lee, and Toutanova}]{devlin_bert_2019}
Jacob Devlin, Ming-Wei Chang, Kenton Lee, and Kristina Toutanova. 2019.
\newblock \href {http://arxiv.org/abs/1810.04805} {{BERT}: {Pre}-training of {Deep} {Bidirectional} {Transformers} for {Language} {Understanding}}.
\newblock ArXiv:1810.04805 [cs].

\bibitem[{Formal et~al.(2021)Formal, Piwowarski, and Clinchant}]{formal_splade_2021}
Thibault Formal, Benjamin Piwowarski, and Stéphane Clinchant. 2021.
\newblock \href {https://doi.org/10.48550/arXiv.2107.05720} {{SPLADE}: {Sparse} {Lexical} and {Expansion} {Model} for {First} {Stage} {Ranking}}.
\newblock ArXiv:2107.05720 [cs].

\bibitem[{G{\"u}nther et~al.(2023)G{\"u}nther, Ong, Mohr, Abdessalem, Abel, Akram, Guzman, Mastrapas, Sturua, Wang et~al.}]{gunther2023jina}
Michael G{\"u}nther, Jackmin Ong, Isabelle Mohr, Alaeddine Abdessalem, Tanguy Abel, Mohammad~Kalim Akram, Susana Guzman, Georgios Mastrapas, Saba Sturua, Bo~Wang, et~al. 2023.
\newblock Jina embeddings 2: 8192-token general-purpose text embeddings for long documents.
\newblock \emph{arXiv preprint arXiv:2310.19923}.

\bibitem[{He et~al.(2018)He, Liu, Liu, Lyu, Zhao, Xiao, Liu, Wang, Wu, She, Liu, Wu, and Wang}]{he2018dureaderchinesemachinereading}
Wei He, Kai Liu, Jing Liu, Yajuan Lyu, Shiqi Zhao, Xinyan Xiao, Yuan Liu, Yizhong Wang, Hua Wu, Qiaoqiao She, Xuan Liu, Tian Wu, and Haifeng Wang. 2018.
\newblock \href {http://arxiv.org/abs/1711.05073} {Dureader: a chinese machine reading comprehension dataset from real-world applications}.

\bibitem[{Karpukhin et~al.(2020)Karpukhin, Oguz, Min, Lewis, Wu, Edunov, Chen, and Yih}]{karpukhin_dense_2020}
Vladimir Karpukhin, Barlas Oguz, Sewon Min, Patrick Lewis, Ledell Wu, Sergey Edunov, Danqi Chen, and Wen-tau Yih. 2020.
\newblock \href {https://doi.org/10.18653/v1/2020.emnlp-main.550} {Dense {Passage} {Retrieval} for {Open}-{Domain} {Question} {Answering}}.
\newblock In \emph{Proceedings of the 2020 {Conference} on {Empirical} {Methods} in {Natural} {Language} {Processing} ({EMNLP})}, pages 6769--6781, Online. Association for Computational Linguistics.

\bibitem[{Khattab and Zaharia(2020)}]{khattab_colbert_2020}
Omar Khattab and Matei Zaharia. 2020.
\newblock \href {https://doi.org/10.48550/arXiv.2004.12832} {{ColBERT}: {Efficient} and {Effective} {Passage} {Search} via {Contextualized} {Late} {Interaction} over {BERT}}.
\newblock ArXiv:2004.12832 [cs].

\bibitem[{Kusupati et~al.(2022)Kusupati, Bhatt, Rege, Wallingford, Sinha, Ramanujan, Howard-Snyder, Chen, Kakade, Jain, and Farhadi}]{kusupati_2022_matryoshka}
Aditya Kusupati, Gantavya Bhatt, Aniket Rege, Matthew Wallingford, Aditya Sinha, Vivek Ramanujan, William Howard-Snyder, Kaifeng Chen, Sham Kakade, Prateek Jain, and Ali Farhadi. 2022.
\newblock \href {https://proceedings.neurips.cc/paper_files/paper/2022/file/c32319f4868da7613d78af9993100e42-Paper-Conference.pdf} {Matryoshka representation learning}.
\newblock In \emph{Advances in Neural Information Processing Systems}, volume~35, pages 30233--30249. Curran Associates, Inc.

\bibitem[{Lawrie et~al.(2023)Lawrie, Yang, Oard, and Mayfield}]{lawrie_neural_2023}
Dawn Lawrie, Eugene Yang, Douglas~W. Oard, and James Mayfield. 2023.
\newblock \href {https://doi.org/10.48550/arXiv.2209.01335} {Neural {Approaches} to {Multilingual} {Information} {Retrieval}}.
\newblock ArXiv:2209.01335 [cs].

\bibitem[{Lee et~al.(2024)Lee, Dai, Ren, Chen, Cer, Cole, Hui, Boratko, Kapadia, Ding, Luan, Duddu, Abrego, Shi, Gupta, Kusupati, Jain, Jonnalagadda, Chang, and Naim}]{lee_gecko_2024}
Jinhyuk Lee, Zhuyun Dai, Xiaoqi Ren, Blair Chen, Daniel Cer, Jeremy~R. Cole, Kai Hui, Michael Boratko, Rajvi Kapadia, Wen Ding, Yi~Luan, Sai Meher~Karthik Duddu, Gustavo~Hernandez Abrego, Weiqiang Shi, Nithi Gupta, Aditya Kusupati, Prateek Jain, Siddhartha~Reddy Jonnalagadda, Ming-Wei Chang, and Iftekhar Naim. 2024.
\newblock \href {http://arxiv.org/abs/2403.20327} {Gecko: {Versatile} {Text} {Embeddings} {Distilled} from {Large} {Language} {Models}}.
\newblock ArXiv:2403.20327 [cs] version: 1.

\bibitem[{Li et~al.(2023)Li, Zhang, Zhang, Long, Xie, and Zhang}]{li2023generaltextembeddingsmultistage}
Zehan Li, Xin Zhang, Yanzhao Zhang, Dingkun Long, Pengjun Xie, and Meishan Zhang. 2023.
\newblock \href {http://arxiv.org/abs/2308.03281} {Towards general text embeddings with multi-stage contrastive learning}.

\bibitem[{Lin et~al.(2021)Lin, Ma, Lin, Yang, Pradeep, and Nogueira}]{Lin_etal_SIGIR2021_Pyserini}
Jimmy Lin, Xueguang Ma, Sheng-Chieh Lin, Jheng-Hong Yang, Ronak Pradeep, and Rodrigo Nogueira. 2021.
\newblock {Pyserini}: A {Python} toolkit for reproducible information retrieval research with sparse and dense representations.
\newblock In \emph{Proceedings of the 44th Annual International ACM SIGIR Conference on Research and Development in Information Retrieval (SIGIR 2021)}, pages 2356--2362.

\bibitem[{Liu et~al.(2019)Liu, Ott, Goyal, Du, Joshi, Chen, Levy, Lewis, Zettlemoyer, and Stoyanov}]{liu2019roberta}
Yinhan Liu, Myle Ott, Naman Goyal, Jingfei Du, Mandar Joshi, Danqi Chen, Omer Levy, Mike Lewis, Luke Zettlemoyer, and Veselin Stoyanov. 2019.
\newblock Roberta: A robustly optimized bert pretraining approach.
\newblock \emph{arXiv preprint arXiv:1907.11692}.

\bibitem[{Louis et~al.(2024)Louis, Saxena, van Dijck, and Spanakis}]{louis_colbert-xm_2024}
Antoine Louis, Vageesh Saxena, Gijs van Dijck, and Gerasimos Spanakis. 2024.
\newblock \href {http://arxiv.org/abs/2402.15059} {{ColBERT}-{XM}: {A} {Modular} {Multi}-{Vector} {Representation} {Model} for {Zero}-{Shot} {Multilingual} {Information} {Retrieval}}.
\newblock ArXiv:2402.15059 [cs].

\bibitem[{Lù(2024)}]{bm25s}
Xing~Han Lù. 2024.
\newblock \href {http://arxiv.org/abs/2407.03618} {Bm25s: Orders of magnitude faster lexical search via eager sparse scoring}.

\bibitem[{Merrick et~al.(2024)Merrick, Xu, Nuti, and Campos}]{merrick2024arcticembedscalableefficientaccurate}
Luke Merrick, Danmei Xu, Gaurav Nuti, and Daniel Campos. 2024.
\newblock \href {http://arxiv.org/abs/2405.05374} {Arctic-embed: Scalable, efficient, and accurate text embedding models}.

\bibitem[{Mohr et~al.(2024)Mohr, Krimmel, Sturua, Akram, Koukounas, G{\"u}nther, Mastrapas, Ravishankar, Mart{\'\i}nez, Wang et~al.}]{mohr2024multi}
Isabelle Mohr, Markus Krimmel, Saba Sturua, Mohammad~Kalim Akram, Andreas Koukounas, Michael G{\"u}nther, Georgios Mastrapas, Vinit Ravishankar, Joan~Fontanals Mart{\'\i}nez, Feng Wang, et~al. 2024.
\newblock Multi-task contrastive learning for 8192-token bilingual text embeddings.
\newblock \emph{arXiv preprint arXiv:2402.17016}.

\bibitem[{Muennighoff et~al.(2023)Muennighoff, Tazi, Magne, and Reimers}]{muennighoff-etal-2023-mteb}
Niklas Muennighoff, Nouamane Tazi, Loic Magne, and Nils Reimers. 2023.
\newblock \href {https://doi.org/10.18653/v1/2023.eacl-main.148} {{MTEB}: Massive text embedding benchmark}.
\newblock In \emph{Proceedings of the 17th Conference of the European Chapter of the Association for Computational Linguistics}, pages 2014--2037, Dubrovnik, Croatia. Association for Computational Linguistics.

\bibitem[{Nair et~al.(2022)Nair, Yang, Lawrie, Duh, McNamee, Murray, Mayfield, and Oard}]{ecir2022colbert-x}
Suraj Nair, Eugene Yang, Dawn Lawrie, Kevin Duh, Paul McNamee, Kenton Murray, James Mayfield, and Douglas~W. Oard. 2022.
\newblock \href {https://arxiv.org/abs/2201.08471} {Transfer learning approaches for building cross-language dense retrieval models}.
\newblock In \emph{Proceedings of the 44th European Conference on Information Retrieval (ECIR)}.

\bibitem[{Penedo et~al.(2023)Penedo, Malartic, Hesslow, Cojocaru, Cappelli, Alobeidli, Pannier, Almazrouei, and Launay}]{penedo2023refinedweb}
Guilherme Penedo, Quentin Malartic, Daniel Hesslow, Ruxandra Cojocaru, Alessandro Cappelli, Hamza Alobeidli, Baptiste Pannier, Ebtesam Almazrouei, and Julien Launay. 2023.
\newblock The refinedweb dataset for falcon llm: outperforming curated corpora with web data, and web data only.
\newblock \emph{arXiv preprint arXiv:2306.01116}.

\bibitem[{Santhanam et~al.(2022)Santhanam, Khattab, Saad-Falcon, Potts, and Zaharia}]{santhanam_colbertv2_2022}
Keshav Santhanam, Omar Khattab, Jon Saad-Falcon, Christopher Potts, and Matei Zaharia. 2022.
\newblock \href {https://doi.org/10.48550/arXiv.2112.01488} {{ColBERTv2}: {Effective} and {Efficient} {Retrieval} via {Lightweight} {Late} {Interaction}}.
\newblock ArXiv:2112.01488 [cs].

\bibitem[{Su et~al.(2022)Su, Shi, Kasai, Wang, Hu, Ostendorf, Yih, Smith, Zettlemoyer, and Yu}]{su2022one}
Hongjin Su, Weijia Shi, Jungo Kasai, Yizhong Wang, Yushi Hu, Mari Ostendorf, Wen-tau Yih, Noah~A Smith, Luke Zettlemoyer, and Tao Yu. 2022.
\newblock One embedder, any task: Instruction-finetuned text embeddings.
\newblock \emph{arXiv preprint arXiv:2212.09741}.

\bibitem[{Su et~al.(2023)Su, Lu, Pan, Murtadha, Wen, and Liu}]{su2023roformerenhancedtransformerrotary}
Jianlin Su, Yu~Lu, Shengfeng Pan, Ahmed Murtadha, Bo~Wen, and Yunfeng Liu. 2023.
\newblock \href {http://arxiv.org/abs/2104.09864} {Roformer: Enhanced transformer with rotary position embedding}.

\bibitem[{Wang et~al.(2023)Wang, Yang, Huang, Yang, Majumder, and Wei}]{wang2023improving}
Liang Wang, Nan Yang, Xiaolong Huang, Linjun Yang, Rangan Majumder, and Furu Wei. 2023.
\newblock Improving text embeddings with large language models.
\newblock \emph{arXiv preprint arXiv:2401.00368}.

\bibitem[{Wang et~al.(2024)Wang, Yang, Huang, Yang, Majumder, and Wei}]{wang2024multilingual}
Liang Wang, Nan Yang, Xiaolong Huang, Linjun Yang, Rangan Majumder, and Furu Wei. 2024.
\newblock Multilingual e5 text embeddings: A technical report.
\newblock \emph{arXiv preprint arXiv:2402.05672}.

\bibitem[{Xiong et~al.(2020)Xiong, Xiong, Li, Tang, Liu, Bennett, Ahmed, and Overwijk}]{xiong_approximate_2020}
Lee Xiong, Chenyan Xiong, Ye~Li, Kwok-Fung Tang, Jialin Liu, Paul Bennett, Junaid Ahmed, and Arnold Overwijk. 2020.
\newblock \href {https://doi.org/10.48550/arXiv.2007.00808} {Approximate {Nearest} {Neighbor} {Negative} {Contrastive} {Learning} for {Dense} {Text} {Retrieval}}.
\newblock ArXiv:2007.00808 [cs].

\bibitem[{Yang et~al.(2024)Yang, Lawrie, and Mayfield}]{yang_distillation_2024}
Eugene Yang, Dawn Lawrie, and James Mayfield. 2024.
\newblock \href {https://doi.org/10.1145/3626772.3657955} {Distillation for {Multilingual} {Information} {Retrieval}}.
\newblock In \emph{Proceedings of the 47th {International} {ACM} {SIGIR} {Conference} on {Research} and {Development} in {Information} {Retrieval}}, pages 2368--2373.
\newblock ArXiv:2405.00977 [cs].

\bibitem[{Zhang et~al.(2021)Zhang, Ma, Shi, and Lin}]{mrtydi}
Xinyu Zhang, Xueguang Ma, Peng Shi, and Jimmy Lin. 2021.
\newblock {Mr. TyDi}: A multi-lingual benchmark for dense retrieval.
\newblock \emph{arXiv:2108.08787}.

\bibitem[{Zhang et~al.(2023{\natexlab{a}})Zhang, Ogueji, Ma, and Lin}]{zhang_2023_towards}
Xinyu Zhang, Kelechi Ogueji, Xueguang Ma, and Jimmy Lin. 2023{\natexlab{a}}.
\newblock \href {https://doi.org/10.1145/3613447} {Toward best practices for training multilingual dense retrieval models}.
\newblock \emph{ACM Trans. Inf. Syst.}, 42(2).

\bibitem[{Zhang et~al.(2023{\natexlab{b}})Zhang, Thakur, Ogundepo, Kamalloo, Alfonso-Hermelo, Li, Liu, Rezagholizadeh, and Lin}]{zhang2023miracl}
Xinyu Zhang, Nandan Thakur, Odunayo Ogundepo, Ehsan Kamalloo, David Alfonso-Hermelo, Xiaoguang Li, Qun Liu, Mehdi Rezagholizadeh, and Jimmy Lin. 2023{\natexlab{b}}.
\newblock Miracl: A multilingual retrieval dataset covering 18 diverse languages.
\newblock \emph{Transactions of the Association for Computational Linguistics}, 11:1114--1131.

\end{thebibliography}
\bibliographystyle{acl_natbib}
\clearpage
\appendix



\end{document}